\documentclass{aa}  

\usepackage{graphicx}
\usepackage{txfonts}
\def\targ{IRAS~21078$+$5211}
\def\wat{H$_{2}$O}
\def\meth{CH$_{3}$OH}
\def\nh3{NH$_{3}$}
\def\kms{km~s$^{-1}$}

\def\Vlsr{$V_{\rm LSR}$}
\def\Jyb{Jy~beam$^{-1}$}

\def\G24{G24.78$+$0.08}

\def\tbo {$T_{\rm{b}}\Delta\Omega$}

\def\dvi {$\Delta V_{\rm{i}}$}

\def\code {FRTM code}

\newcommand{\ms}{$M_{\odot}$}
\newcommand{\ls}{$L_{\odot}$}

\newcommand{\pas}{$\rlap{.}^{\prime\prime}$}

\newcommand{\degree}{$^{\circ}$}

\usepackage{xcolor}

\begin{document}

\title{The magnetic field of a magnetohydrodynamic disk wind: Water maser observations and simulations}
   
  \titlerunning{Magnetic field in a MHD disk wind}

   \author{L. Moscadelli \inst{1}
          \and
           A. Oliva \inst{2,3,4}
          \and
          G. Surcis \inst{5}
          \and
          A. Sanna \inst{5}
          \and
          M.~T. Beltr{\'a}n \inst{1}  
          \and
          R. Kuiper\inst{6}
          }

   \institute{INAF-Osservatorio Astrofisico di Arcetri, Largo E. Fermi 5, I-50125, Firenze, Italy \\
              \email{luca.moscadelli@inaf.it}
         \and
         Observatoire de Genève, Université de Genève, Chemin Pegasi 51, CH-1290 Versoix, Switzerland
         \and
         Zentrum f{\"u}r Astronomie der Universit{\"a}t Heidelberg, Institut f{\"u}r Theoretische Astrophysik, Albert-Ueberle-Stra{\ss}e 2, 69120 Heidelberg, Germany
        \and
        Space Research Center (CINESPA), School of Physics, University of Costa Rica, 11501 San Jos\'e, Costa Rica
         \and
         INAF - Osservatorio Astronomico di Cagliari, Via della Scienza 5, 09047 Selargius (CA), Italy
         \and
         Faculty of Physics, University of Duisburg-Essen, Lotharstra{\ss}e 1, D-47057 Duisburg, Germany
             }


 
  \abstract
   {Although \(\)star-formation models predict that the magnetic field plays an important role in regulating disk-mediated accretion and launching and collimating protostellar jets, observations of the magnetic field  close enough  (within a few 100~au) to the forming stars are still sparse.}
   {Our goal is to measure and model the magnetic field distribution in the disk wind of the young stellar object (YSO) \targ.}
   {We performed sensitive global Very Long Baseline Interferometry observations of the polarized emission of the 22~GHz water masers tracing individual streamlines of the magnetohydrodynamic (MHD) disk wind in \targ. Our resistive-radiative-gravito-MHD simulations of a jet around a forming massive star are able to closely reproduce the observed maser kinematics in the inner jet cavity.}
   {We measure a weak level of 0.3\%--3.2\% of linear and circular polarization in 24 and 8 water masers, respectively. The detected polarized masers sample the direction and the strength of the magnetic field along five distinct streamlines within the inner 100~au region of the disk wind. Along the four streamlines at smaller radii from the jet axis ($\le$~25~au), the sky-projected direction of the magnetic field forms, in most cases, a small offset angle of $\le$30\degree\  with the tangent to the streamline. Along the stream at larger radii (50--100~au), the magnetic field is sampled at only three separated positions, and  it is found to be approximately perpendicular to the streamline tangent at heights of \ $\approx$10~and~40~au, and parallel to the tangent at \ $\approx$70~au. According to our simulations, the magnetic field lines should coincide with the flow streamlines in the inner jet cavity. The small tilt in the magnetic field direction observed along the inner streams can be well explained by Faraday rotation, assuming a realistic low level of ionization for the molecular shell of the jet of namely  $\sim$10$^{-2}$ . The magnetic field amplitudes measured from maser circular polarization are all within a relatively small range of \ 100--700~mG, which is in good agreement with the simulation results and consistent with reduced magnetic diffusivity in the jet cavity owing to efficient shock ionization.}
   {By comparing observations achieving sub-au linear resolution with source-specific simulations, this work presents a very detailed study of the gas kinematics and magnetic field configuration in the MHD disk wind associated with the YSO \targ. The close correspondence between flow streamlines and magnetic field lines together with the relatively high strength of the magnetic field indicate that the magnetic field has a dominant role in the launch and collimation of the YSO jet.}

  \keywords{ISM: jets and outflows -- ISM: kinematics and dynamics -- Stars: formation -- Masers -- Techniques: interferometric}

   \maketitle
%

\section{Introduction}

Dust polarization observations are routinely conducted in order to study the magnetic field in both low- and high-mass star forming regions (on scales of $\sim$~0.1~pc), as they allow us to determine
the field orientation and to assess the relative magnitude of the ordered and turbulent components of the field \citep[e.g.,][]{Gir06,Gir09}. Dust polarization
measurements can also be used to indirectly estimate the magnitude of the magnetic field component on the sky plane based on the \citet{Chan53}
method, which quantifies how much the magnetic field is disturbed by the ``turbulent'' gas motions.  Nevertheless, different effects can contribute to dust
polarization especially at the high-density regimes of circumstellar regions on scales of $\lesssim$~1000~au, such as radiative grain alignment \citep{Laz07} or dust scattering \citep{Kat15}.
For instance,  the latter effect becomes increasingly important when grains grow to sizes that are comparable to the wavelength of the millimeter radiation \citep[$>50\,\mu$m;][]{Kat17},
 as expected inside dusty disks, where pebble accretion happens. This topic has been  widely discussed in the recent literature
\citep[e.g.,][]{Bert17,Alv18,Gir18,Bac18,Hul18,Den19}. Linear polarization of thermal lines such as CO and SiO, also known as the Goldreich-Kylafis effect \citep{Gol81},
can also be used to probe magnetic fields in the plane of the sky, although only marginal detections have been reported so far \citep[e.g.,][]{Ste20,Tea21}.

A more direct measurement of magnetic field strengths can be obtained through Zeeman splitting in the circular polarization of suitable lines, such as the paramagnetic
radical CN \citep[e.g.,][]{Fal08}. The circularly polarized signal produced by Zeeman splitting depends directly on the magnetic field strength along the line of sight (LOS).
However, the detection of Zeeman splitting through thermal line emission coming from the circumstellar regions is challenging because of the high sensitivity and high spectral
resolution required to detect the low level of circular polarization expected (of the order of $1\%$). As a consequence,  Zeeman splitting has so far only been detected at
optical wavelengths toward the innermost region of the accretion disk of FU Orionis \citep{Don05} and, despite significant efforts, no detections have yet been reported at
(sub)millimeter wavelengths for scales of 100\,au or less   \citep[e.g., TW Hya, AS 209:][]{Vle19, Har21}.

In the last two decades, the magnetic field around massive young stellar objects (YSOs) has also been targeted by observing polarized maser emissions with the Very Long Baseline Interferometry (VLBI) technique \citep[e.g.,][]{Vle12, Sur22}. Specific targets are the intense 6.7\,GHz \meth\ and 22\,GHz \wat\ maser emissions, which
have been routinely observed in recent decades for gas kinematics studies within 10--1000\,au of high-mass YSOs \citep[e.g.,][]{Mos07,San10a,Mos11a,God11a}.
Because of the low level ($\lesssim$1\%) of maser polarization fraction, with the typical  spectral  line VLBI sensitivity of \ $\gtrsim$10~mJy per 0.1~\kms, only the most
intense ($\ge$~10--100~Jy) sources yield a significant number of independent polarization measurements ($>10$).  Despite this limitation, in a few
objects, VLBI of \wat\ masers has succeeded in adequately mapping the magnetic field over the disk--jet region within a few 100\,au of the YSO. The first case studied
is that of the massive star-forming region W75N(B) \citep{Sur11,Sur14,Sur23}, where water masers revealed the different morphologies of the magnetic field associated
with two YSOs, named VLA\,1 and VLA\,2, which are thought to be in different evolutionary stages. A second case study of H$_2$O maser polarization is the high-mass
YSO W3(\wat), where \cite{God17} found that the magnetic field probed by 22\,GHz \wat\ masers is oriented along the axis of a synchrotron jet detected in the
radio continuum at scales of 1000\,au; the same magnetic field direction is instead misaligned with respect to the maser proper motions on scales of\ 10--100\,au.
On the other hand, the 6.7\,GHz \meth\ masers are generally distributed at distances of greater than 300\,au from the YSO, tracing magnetic fields in the outer disk
layers and likely the interaction between circumstellar envelopes and outflowing gas \citep{Sur15,Sur19,Sur22}. A case study in this respect is that of the O-type
YSO named G023.01$-$00.41 \citep{San15}, where the magnetic field appears to be correlated with the velocity field of the gas close to the disk midplane (at radii
\ $\ge$~600~au), and to be  more turbulent closer to the jet axis above this plane.

From a theoretical perspective, several recent numerical studies have built on the classical description of disk winds \citep[DWs; see, e.g.,][]{Bla82,Pel92} and magnetohydrodynamic (MHD) simulations in the ideal approximation \citep[see, e.g.,][]{Bane07,Sei12} to widen our understanding of how magnetic fields drive outflows during the formation of massive stars. \citet{Koe18}, \citet{Matsu17}, and \citet{Machi20} performed MHD simulations with Ohmic dissipation as a nonideal effect and approximations to thermodynamics (an isothermal equation of state in the first study and a barotropic equation of state in the others). While \citet{Koe18} find a magneto-centrifugally launched jet, \citet{Matsu17} and \citet{Machi20} do not find an outflow in any of the cases studied. \citet{Migno21} and \citet{Com22} considered ambipolar diffusion as a nonideal effect in their 3D MHD simulations but no Ohmic dissipation, and both studies included a diffusive treatment for radiation transport of the dust and gas emission. These latter studies found indications of magnetically driven outflows of typical velocities of the order of 30~\kms, and magnetic fields strengths that are below $\sim$0.1~G throughout. These latter authors used the RAMSES code with an adaptive mesh refinement grid with a minimum cell size of 5~au. 
\citet{Oli23a,Oli23b} performed MHD simulations with the Pluto code  \citep{Mig07} including the effects of radiation transport and Ohmic dissipation. Because of their axisymmetric grid setup in spherical coordinates, the authors were able to reach high resolution (up to 0.03~au) in the innermost 100~au, allowing for a detailed study of the launch and propagation of a magneto-centrifugal jet. These latter studies also explored the effects of the initial conditions on the evolution of the disk--jet system and the YSO.

In \citet{Mos22}, we used one of the simulations described in \citet{Oli23a,Oli23b} to perform an order-of-magnitude comparison of the kinematics of the jet and the water masers observed in \targ. Our first results showed that the simulations are able to reproduce a helical fast flow close to the YSO, the observed size of the disk, the mass of the YSO, and the overall structure of the slower molecular flow at large scales.

\begin{figure*}%
\includegraphics[width=\textwidth]{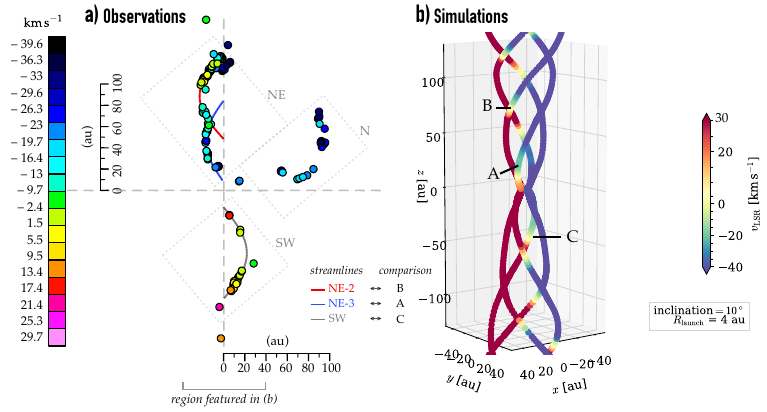}
\caption{{\bf 22~GHz water masers in \targ\ and a 3D view of the DW streamlines that best reproduce the maser kinematics.} \ (a)~Water masers found in the global VLBI observation in 2020 October. Colored dots give the relative positions of the 22~GHz water masers, with colors denoting the local-standard-of-rest velocity of the masers, \Vlsr. We note that the observed positions have been rotated 50\degree \ clockwise so as to align the jet axis with the vertical axis (and the disk axis with the horizontal axis). 
The black dotted rectangles encompass the three regions, to the N, NE, and SW, where the maser emission concentrates. The black, red, and blue curves denote the sinusoids drawn by the masers belonging to the SW, NE-2, and NE-3 streams, respectively.
\ (b)~Simulation of the DW kinematics. The colored curves denote four DW streamlines departing from the launch radius of 4~au at 90\degree\ azimuthal separation from one another, as computed with the simulation that best reproduces the geometry and kinematics of the maser patterns (see Sect.~\ref{simu-fit}). Colors show \Vlsr\ using a scale similar to that of the observations. The simulations reproduce only the inner maser streams, within a radius of 40~au. The streamlines labeled ``A'', ``B'', and ``C'' are those that best reproduce the NE-3 (blue sinusoid), NE-2 (red sinusoid), and SW (black sinusoid) maser streams, respectively. We note that the NE-2 stream is, strictly speaking, better reproduced using a slightly larger launch radius of 4.5~au (see Sect.~\ref{simu-fit}). In agreement with previous observational constraints (see Sect.~\ref{Res-1}), \Vlsr\ is calculated for a LOS inclination of the jet axis of 10\degree.}
\label{glo}
\end{figure*}

\targ \ is a star forming region of high bolometric luminosity, 
\ 5$\, \times \,$10$^{3}$~\ls \ \citep{Mos16} at a distance of \ 1.63$\pm$0.05~kpc \citep{Xu13}, and harbors a cluster of forming massive stars. In recent years, the main properties (3D orientation, size, kinematics) of the disk--jet system around the most massive YSO $(5.6\pm2$~\ms) in the cluster have been studied at high angular resolutions (1--100~mas) using NOrthern Extended Millimeter Array (NOEMA) molecular line, Jansky Very Large Array (JVLA) continuum (1.3 and 5~cm), and multi-epoch (2010--2011) Very Long Baseline Array (VLBA) 22~GHz water maser observations \citep{Mos16,Mos21}. In October 2020, we reobserved the water masers with sensitive global VLBI observations, and demonstrated that the water masers trace streamlines of gas emerging from the YSO disk and spiralling outward along the jet axis, providing us with a direct view of the velocity field of the YSO's MHD DW \citep{Mos22}. Figure~\ref{glo}a shows the water maser positions around the YSO, mainly found within radii \ $\le$100~au. Close to the disk rotation axis, in the SW and NE regions, the masers trace sinusoidal patterns in the plane of the sky (see Fig.~\ref{glo}a), which are univocal signatures of a spiralling motion along the jet axis. 
At larger separation from the rotation axis, in the N region, the local-standard-of-rest velocity (\Vlsr) of the  maser \ changes linearly with the radial distance,
which indicates that the maser stream co-rotates with its launch point from the disk as predicted by magneto-centrifugal acceleration. Our simulations of a MHD DW around a forming massive star yield streamlines that closely match the maser velocity patterns (see Fig.~\ref{glo}b).

This work reports the configuration of the magnetic field in the DW of \targ,\ as derived from our recent global VLBI polarization observations of the 22~GHz water masers, together with the results of the simulations performed to reproduce both the velocity and magnetic fields traced with the masers.
Section~\ref{obse} describes the observations and the  analysis we performed to derive the magnetic field. In Sect.~\ref{Res}, we show the measurements of the magnetic field, illustrate the initial conditions and fundamental parameters of our simulations, and describe the strategy we employed to select the simulation snapshot that best reproduces the observed data. Section~\ref{discu} discusses the physical characteristics of our best-matching simulation and its limits and corresponding difficulties in reproducing the observations, and compares this simulation with other simulations in our catalog. Finally, we draw conclusions in Sect.~\ref{conclu}.

\section{VLBI polarization observations and magnetic field measurement}

\label{obse}

Descriptions of the global VLBI observations, correlation, and data calibration were reported in \citet{Mos22}. Here, we focus on the aspects relevant to the polarization observations only.  We recorded dual circular polarization signals through four adjacent base bands, each with a width of 16 MHz, and one of them centered at the maser \Vlsr \ of \ $-$6.4~\kms. The maser base band split into a narrower width of 8~MHz was correlated with the European VLBI Network (EVN) software correlator \citep[SFXC,][]{Keip15} at the Joint Institute for VLBI ERIC (JIVE, at Dwingeloo, the Netherlands) with 4096 channels in order to achieve the very high velocity resolution of 0.026~\kms\ needed for revealing circular polarization.
The right and left circularly polarized radio signals received and registered at the antennas were also cross-correlated to generate all four polarization combinations (RR, LL, RL, and LR).

Data were reduced with the Astronomical Image Processing System (\textsc{AIPS}) package following the VLBI spectral line
procedures \citep{San10a}. After applying the parallel-polarization (amplitude and phase) calibration, the delay corrections for the cross-polarization signals and the feed polarization parameters were determined from scans of the polarization calibrator \ J2202$+$4216. The maser polarization angles, or electric vector position angles (EVPAs) of linearly polarized maser emission, were corrected so as to set the polarization angle of 3C48 --- which was observed as the secondary polarization calibrator --- to \ $-$70\degree$\pm$2\degree\ \citep{Per13}. The emission of an intense and compact maser channel was self-calibrated, and the derived (amplitude and phase) corrections were applied to all the maser channels before imaging. We produced images at the full velocity resolution of the four Stokes parameters \ \textit{I}, \textit{Q}, \textit{U,} and \textit{V}, covering the whole emission range of maser detections by \citet{Mos22} of namely  [$-$40, $+$30]~\kms. To save computational time, we mapped a relatively small area extending  \ 0\pas082  \ in both right ascension (RA) $\cos \delta$ \ and declination\ (DEC) around the YSO position. This area includes the strongest maser features (with intensity $\ge$~1.5~\Jyb) distributed inside the NE, SW, and N regions where the water maser streamlines are detected (see Fig.~\ref{glo}a). Given that polarization is detected above maser intensities of $\ge$~0.7~\Jyb (see Table~\ref{pol_fit}), we could have only missed the detection of polarized emission for a single feature \citep[label number \#32 in Supplementary Table~1 of][]{Mos22} outside the mapped area. Using natural weighting, the FWHM major and minor sizes of the beam are\ 0.3~mas and 0.2~mas, respectively, and the beam PA is \ 1.7\degree. In channel maps devoid of emission, the 1$\sigma$ rms noise is 1.6~mJy~beam$^{-1}$, which is close to the expected thermal noise.

\begin {table*}[t!]
\caption []{Fitted parameters of the polarized 22 GHz \wat ~maser features.} 
\begin{center}
\scriptsize
\begin{tabular}{ l c c c c c c c c c c c c c}
\hline
\hline
   1  &  2       &  3     &  4  & 5 &  6   &  7        &  8     &   9  & 10    &  11     &  12   & 13 & 14   \\
Maser& RA\tablefootmark{a}&Dec\tablefootmark{a}& Maser & Peak flux & $V_{\rm{LSR}}$& $P_{\rm{l}}$ &  $\chi$   & $\Delta V_{\rm{i}}\tablefootmark{b}$ & $T_{\rm{b}}\Delta\Omega\tablefootmark{b}$& $P_{\rm{V}}$ & $B_{\rm LOS}$  &$\theta$ & prob. \\
ID     &  offset        &  offset     &  Stream  & Density(I)&           &              &            &               &                      &              &                      &  &  $B_{||}$  \\ 
     &  (mas)         &  (mas)      &    & (Jy/beam) &  (km/s)   &  (\%)        &   (\degree)     &   (km/s)   & (log K sr)           &   ($\%$)     &  (mG)               &(\degree)  & (\%)      \\ 
\hline
 1  & 0               & 0     & $-$         & $53.066$  & -21.6     &              $0.57\pm0.01$ & $+46\pm6$  & $3.2^{+0.1}_{-0.5}$ & $10.09^{+0.2}_{-0.2}$& $-$  & $-$          &$52^{+6}_{-42}$  & 94 \\ 
 2  & $+60.42\pm0.03$ & $+27.87\pm0.03$ & NE-2  & $14.326$  & -3.0      &              $0.3\pm0.1$   & $+36\pm47$ & $3.3^{+0.2}_{-0.2}$ & $8.7^{+0.6}_{-1.2}$ & $-$   & $-$          &$68^{+8}_{-44}$  &  60 \\ 
 3  & $-14.93\pm0.03$ & $+51.87\pm0.03$ & N  & $14.017$  & -24.0     &              $1.09\pm0.23$ & $+29\pm2$  & $3.3^{+0.1}_{-0.4}$ & $9.4^{+0.6}_{-1.6}$ & $-$   & $-$          &$67^{+12}_{-39}$  & 53 \\ 
 4  & $-1.55\pm0.03$  & $+58.23\pm0.03$ &  N  & $10.457$  & -28.5     &              $0.4\pm0.1$   & $+40\pm20$ & $2.8^{+0.2}_{-0.2}$  & $9.1^{+0.7}_{-6}$  & $0.4\pm0.1 $ & $-160\pm80$  & $62^{+2}_{-44}$  & 80 \\ 
 5  & $-42.37\pm0.03$ & $-36.16\pm0.03$ &  SW  & $7.082$   & +0.4      &              $0.5\pm0.2$   & $+22\pm14$ & $3.3^{+0.2}_{-0.2}$ & $9.1^{+0.5}_{-1.5}$ & $-$   & $-$          &$66^{+9}_{-43}$  & 62 \\ 
 6  & $+56.89\pm0.03$ & $+18.87\pm0.03$ &  NE-2 & $6.565$   & -1.0      &              $0.3\pm0.1$   & $+56\pm15$ & $2.9^{+0.1}_{-0.2}$ & $8.6^{+0.8}_{-0.9}$ & $1.4\pm0.2 $ & $-457\pm144$ &$67^{+8}_{-44}$   &  62 \\ 
 7  & $+23.06\pm0.03$ & $-4.24\pm0.03$  &  NE-3 & $6.502$   & -15.2     &              $0.41\pm0.06$ & $+10\pm4$  & $3.8^{+0.1}_{-0.3}$ & $9.0^{+0.5}_{-0.5}$ & $0.7\pm0.2 $ & $+319\pm188$ &$67^{+6}_{-42}$   & 63 \\ 
8   & $-1.29\pm0.03$  & $+59.10\pm0.03$ &  N & $5.543$   & -30.2     &              $0.3\pm0.1$   & $+38\pm27$ & $3.8^{+0.1}_{-0.3}$  & $8.7^{+0.7}_{-0.2}$ & $-$    & $-$         &$68^{+4}_{-44}$  & 65  \\ 
10  & $+33.59\pm0.03$ & $+3.34\pm0.03$  &  NE-3 & $4.866$   & -15.1     &              $0.45\pm0.19$ & $+71\pm8$  & $2.9^{+0.1}_{-0.3}$ & $9.1^{+0.7}_{-1.4}$ & $1.3\pm0.2 $ & $+441\pm154$ &$62^{+8}_{-45}$   & 72 \\  
11  & $+58.57\pm0.03$ & $+32.40\pm0.03$ & NE-1  & $4.812$   & -19.8     &              $0.39\pm0.05$ & $+32\pm6$  & $3.2^{+0.3}_{-0.1}$ & $8.9^{+0.3}_{-0.7}$ & $-$   & $-$          &$68^{+15}_{-34}$      &  43     \\ 
12  & $-14.78\pm0.03$ & $+52.52\pm0.03$ & N &  $4.395$   & -24.8     &              $1.35\pm0.33$ & $+11\pm2$  & $3.3^{+0.1}_{-0.6}$ & $8.4^{+0.5}_{-1.7}$ & $-$   & $-$          &$90^{+22}_{-22}$      &  0    \\  
13  & $-43.75\pm0.03$ & $-40.90\pm0.03$ &  SW & $4.273$   & +5.0      &              $0.3\pm0.1$   & $+54\pm6$  & $3.0^{+0.1}_{-0.4}$ & $8.5^{+0.8}_{-0.3}$ & $1.4\pm0.1 $ & $-497\pm211$ &$70^{+5}_{-46}$   &  61\\  
14  & $+60.79\pm0.03$ & $+28.54\pm0.03$ &  NE-2 & $4.072$   & -3.2      &              $0.6\pm0.1$   & $+39\pm5$  & $3.6^{+0.2}_{-0.2}$ & $9.2^{+0.3}_{-0.7}$ & $-$   & $-$          &$64^{+6}_{-41}$   &  68 \\   
16  & $+61.14\pm0.03$ & $+31.34\pm0.03$ & NE-2  & $3.397$   & +1.5      &              $3.2\pm0.2$   & $+27\pm4$  & $2.3^{+0.1}_{-0.2}$ & $10.6^{+0.2}_{-0.1}$& $-$   & $-$          &$42^{+11}_{-16}$  & 100 \\  
17  & $-41.84\pm0.03$ & $-34.58\pm0.03$ & SW  & $3.387$   & -0.1      &              $0.4\pm0.1$   & $+87\pm23$ & $2.8^{+0.1}_{-0.3}$ & $8.8^{+0.4}_{-1.4}$ & $-$   & $-$          &$69^{+15}_{-35}$      &  42    \\   
18  & $+33.76\pm0.03$ & $+3.88\pm0.03$  & NE-3 &  $3.206$   & -15.4     &              $0.49\pm0.08$ & $+58\pm8$  & $3.3^{+0.2}_{-0.2}$ & $8.9^{+0.4}_{-1.2}$ & $1.0\pm0.2 $  & $+384\pm242$ &$69^{+14}_{-35}$        & 43  \\ 
19  & $-11.28\pm0.03$ & $+22.53\pm0.03$ & N  & $3.177$   & -22.7     &              $0.61\pm0.07$ & $-22\pm7$  & $3.6^{+0.1}_{-0.5}$ & $8.8^{+0.4}_{-1.1}$ & $-$   & $-$          &$78^{+11}_{-18}$          & 0 \\ 
20  & $+53.03\pm0.03$ & $+19.32\pm0.03$ & NE-1 &  $2.485$   & -11.7     &              $0.6\pm0.1$   & $-30\pm9$  & $2.5^{+0.1}_{-0.3}$ & $9.2^{+0.7}_{-1.4}$ & $1.7\pm0.2 $ & $+496\pm308$ &$64^{+10}_{-41}$  & 63 \\ 
24  & $-44.37\pm0.03$ & $-39.19\pm0.03$ & SW &  $2.072$   & +4.3      &              $0.6\pm0.1$   & $+32\pm5$  & $3.3^{+0.1}_{-0.3}$ & $9.1^{+0.4}_{-0.8}$ & $-$   & $-$          &$67^{+9}_{-41}$   &  58 \\
25  & $+25.54\pm0.03$ & $-3.08\pm0.03$  & NE-3  & $2.062$   & -16.2     &              $-$           & $-$        & $-$                 & $-$                 & $1.8\pm0.4\tablefootmark{c}$ & $+715\pm499\tablefootmark{c}$ &$-$  & $-$ \\ 
27  & $+60.50\pm0.03$ & $+38.38\pm0.03$ & NE-2  & $1.926$   & -6.2      &              $0.8\pm0.1$   & $+34\pm12$ & $3.5^{+0.1}_{-0.4}$ & $8.8^{+0.6}_{-1.3}$ & $-$   & $-$          &$86^{+4}_{-39}$       &    19   \\ 
30  & $+62.36\pm0.03$ & $+37.98\pm0.03$ & NE-2  & $1.631$   & -10.8     &              $2.2\pm0.2$   & $-15\pm4$  & $3.6^{+0.3}_{-1.0}$ & $9.7^{+0.8}_{-1.9}$ & $-$   & $-$          &$66^{+4}_{-38}$   & 64  \\ 
31  & $-43.54\pm0.03$ & $-40.37\pm0.03$ & SW  & $1.537$   & +5.2      &              $0.7\pm0.1$   & $+66\pm22$ & $3.2^{+0.2}_{-0.5}$ & $9.6^{+0.5}_{-0.6}$ & $-$   & $-$          &$61^{+1}_{-44}$   &  84 \\  
36  & $+61.04\pm0.03$ & $+38.18\pm0.03$ & NE-2   & $0.954$   & -8.1      &              $1.4\pm0.2$   & $+38\pm9$  & $2.9^{+0.3}_{-0.2}$ & $8.6^{+0.4}_{-1.9}$ & $-$   & $-$          &$90^{+20}_{-20}$       &  0   \\   
40  & $+59.72\pm0.03$ & $+27.76\pm0.03$ & NE-2  & $0.756$   & +0.4      &              $1.8\pm0.2$   & $+58\pm10$ & $2.4^{+0.1}_{-0.2}$ & $10.6^{+0.1}_{-0.2}$& $-$   & $-$          &$48^{+1}_{-31}$   &  100 \\  
\hline
\end{tabular} \end{center}
\tablefoot{Columns~1--3, 5,~and~6 report the label number, the position (RA and DEC) relative to the reference feature, the intensity, and the \Vlsr \ of the water masers from \citet[][see Supplementary Table~1]{Mos22}.
Column~4 reports the streamline to which the maser belongs. Columns~7~and~11 list the measured level of linear and circular polarization, respectively, and Col.~8 the PA of the linear polarization. Columns~9, 10, 12,~and~13 provide the model-derived values of the intrinsic maser line width, $\Delta V_{\rm{i}}$, the product of the maser brightness temperature and beaming angle, $T_{\rm{b}}\Delta\Omega$, the LOS component of the magnetic field, and the angle, $\theta$, between the magnetic field and the LOS, respectively. Column~14 reports the estimated probability that the magnetic field is oriented parallel to the linear polarization. \\
\tablefoottext{a}{The reference position is \ RA(2000) = $21^{\rm{h}}09^{\rm{m}}21^{\rm{s}}\!.7099$ \ and \ 
DEC(2000) = $52^{\circ}22'37''\!\!.001$.}
\tablefoottext{b}{The best-fitting results obtained using a model based on the radiative transfer theory of \wat ~masers 
for $\Gamma+\Gamma_{\nu}=1~\rm{s^{-1}}$ \citep{Sur11}, where \ $\Gamma$ \ is the maser decay rate and \ $\Gamma_{\nu}$ \ the cross-relaxation rate.} 
 \ \tablefoottext{c}{In the fitting model, we include the values $\langle$\tbo$\rangle=9.9\times10^8$~K~sr and 
$\langle$\dvi$\rangle=3.3$~\kms ~that best fit the total intensity emission. }
}
\label{pol_fit}
\end{table*}

We produced the linearly polarized intensity ($POLC=\sqrt{Q^{2}+U^{2}-\sigma_{P}^2}$) and 
polarization angle ($POLA=1/2\times~\arctan (U/Q)$) images by combining the 
Stokes \textit{Q} and \textit{U} images. The polarized intensity was corrected according to the
noise of 
$\sigma_{P}=\sqrt{[(Q \times \sigma_{Q})^2+(U \times \sigma_{U})^2]/(Q^{2}+U^{2})}$, where
$\sigma_{Q}$ and $\sigma_{U}$ are the noise of the corresponding Stokes images. 
The formal error on $POLA$ due to the thermal noise is given by 
$\sigma_{POLA}=0.5 ~(\sigma_{P}/POLC)$
\citep{War74}. Subsequently,
for each maser detected by \citet{Mos22} that showed polarized emission, we measured
the mean linear polarization fraction ($P_{\rm{l}}$) and the mean linear polarization
angle ($\chi$). This was done by considering only the consecutive channels
(more than two) across the total intensity spectrum for which the polarized intensity is
$\geq4~\rm{rms}$. Polarimetric data
were analyzed using the full radiative transfer method (FRTM) code developed by 
\citet{Vle06} and based on the model for unsaturated 22~GHz \wat\ masers of \citet{Ned92}. In this code, the total intensity and linear polarization 
spectra are modeled by gridding the intrinsic maser line width (\dvi) between 0.4 and 4.5~\kms in
steps of 0.025~\kms using a least-square fitting routine ($\chi^2$-model) with an upper limit on the product of the emerging maser  brightness temperature and beaming angle of \tbo$~=10^{11}$~K~sr. This allows us to determine the 
values of \tbo ~and \dvi ~that produce the best-fit models to the linearly polarized emission
of our masers. It should be noted that in the case of saturated masers, the \code ~provides only 
a lower limit for \tbo ~and an upper limit for \dvi. In addition, from the maser polarization theory, it is possible to estimate
the angle between the maser propagation direction and 
the magnetic field ($\theta$) from \tbo ~and $P_{\rm{l}}$ \citep[see, e.g.,][]{Ned90,Ned92}. 
This $\theta$ angle is important because it allows us to solve
the 90\degree ~ambiguity of the magnetic field orientation with respect to the linear polarization vectors. Indeed, if $\theta>\theta_{\rm{crit}}=55$\degree, the magnetic field appears to be perpendicular to the linear polarization vectors; otherwise, it appears parallel \citep{Gol73}. 
To determine the relative orientation of the magnetic field, we consider the
associated errors $\varepsilon^{\rm{\pm}}$ of $\theta$, which are determined by analyzing the
probability distribution function of the full radiative transfer $\chi^2$-model fits, as we also did for \tbo ~and \dvi,  where the plus and minus signs indicate the positive and negative errors, respectively. In particular, considering that $\theta^{\rm{\pm}}=\theta+\varepsilon^{\rm{\pm}}$, if $|\theta^{\rm{+}}-55$\degree$|<|\theta^{\rm{-}}-55$\degree$|$ the magnetic field is more likely parallel to the linear polarization vectors, while if 
$|\theta^{\rm{+}}-55$\degree$|>|\theta^{\rm{-}}-55$\degree$|$ the magnetic field is more likely 
perpendicular. In the last column of Table~\ref{pol_fit}, we report the probability that the magnetic field is parallel to the linear polarization vector.\\
\indent In case where a maser also showed circular polarization ($P_{\rm{V}}$), we measured the Zeeman splitting and estimated the magnetic field strength along the LOS ($B_{\rm LOS}$). For this purpose, we fitted the observed \textit{I} and \textit{V} spectra using the \textit{I} and \textit{V} models produced in the \code ~from the best estimates of \tbo ~and \dvi. However, due to the typical weak circularly polarized emission of \wat ~masers ($<1\%$), it is important to determine whether or not 
the circularly polarized emission is real by measuring the self noise ($\sigma_{\rm{s.-n.}}$) produced by the 
masers in their channels. This noise becomes important when the power contributed by the astronomical maser is 
a significant portion of the total received power \citep{Sau12}. Therefore, we only consider a detection of 
circularly polarized emission to be real when the \textit{V} peak intensity of a maser feature is both $>3~\rm{rms}$ and $>3~\sigma_{\rm{s.-n.}}$.

\begin{figure*}%
\sidecaption
\includegraphics[width=0.7\textwidth]{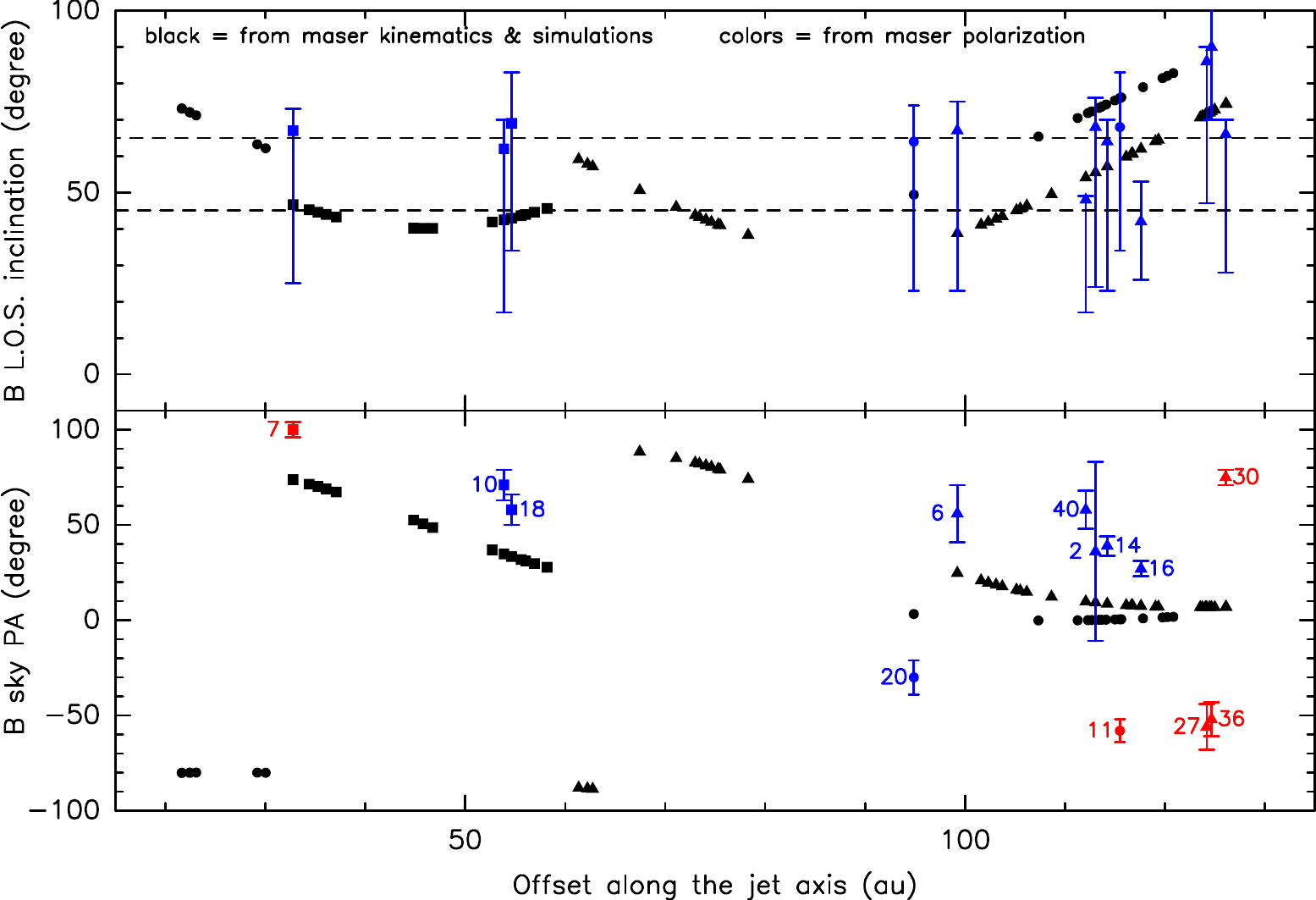}
\caption{{\bf Geometry of the magnetic field vector in the NE region.} Black dots, triangles, and squares indicate the sky-projected orientation (lower panel) and inclination (upper panel) of the magnetic field in relation to the water masers belonging to the NE-1, NE-2, and NE-3 streamlines, respectively, assuming that the magnetic field lines coincide with the maser streamlines. The inclination angle of the jet axis with the plane of the sky is taken to be  \ 15\degree. In the upper panel, the blue symbols with error bars give the value and corresponding uncertainty in the magnetic field inclination from the maser polarization fit (see Table~\ref{pol_fit}). The two black dashed lines mark the inclinations of 45\degree \ and 65\degree.
In the lower panel, colored symbols with error bars denote the determined magnetic field orientation, using blue or red colors if the LOS inclination of the magnetic field is lower or higher than 55\degree, respectively. Masers are identified using the same label numbers as in Table~\ref{pol_fit}.}
\label{NE_geo}
\end{figure*}

\begin{figure}
\centering
\includegraphics[width=0.35\textwidth]{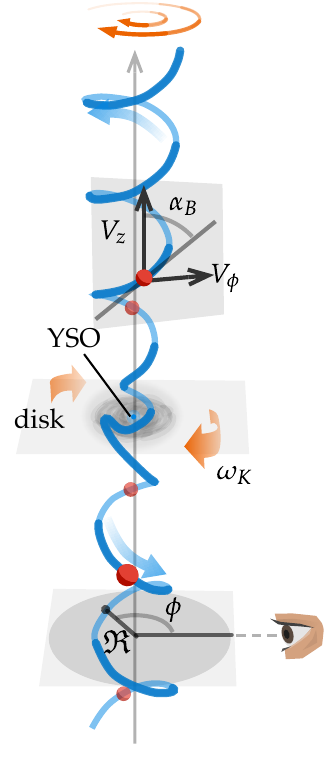}
\caption{{\bf Sketch of a helical motion.} The main geometrical and kinematical parameters are indicated: the radius \ $ \mathfrak{R}$ \ and angle \ $\phi$ \ of rotation; the axial and azimuthal velocity components, $V_{\rm z}$ \ and \ $V_{\phi}$, respectively; and the helix angle \ $\alpha_B$.}
\label{hel-mot}
\end{figure}

\begin{figure}%
\includegraphics[width=0.5\textwidth]{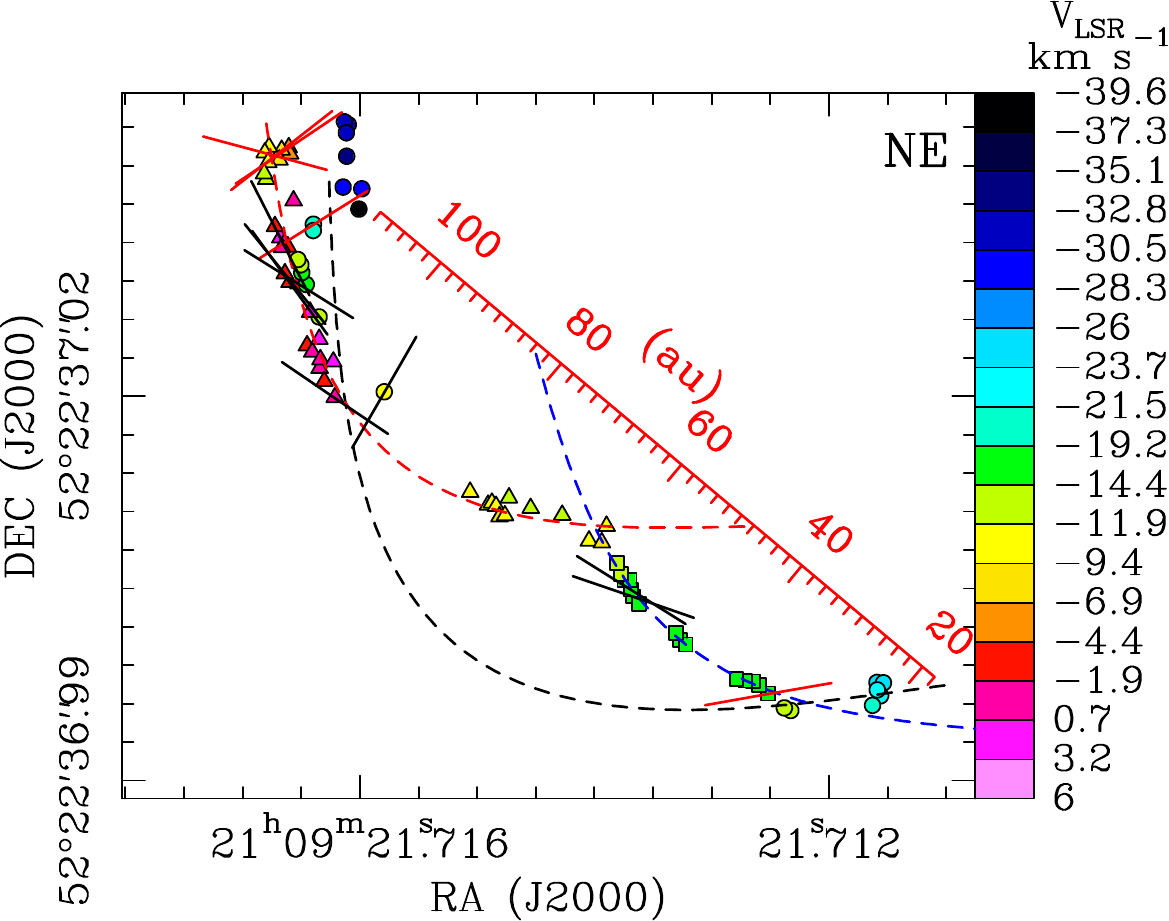}
\caption{{\bf Magnetic field orientation in the NE region.} Colored dots, triangles, and squares give the absolute positions of the 22~GHz water masers belonging to three distinct DW streamlines, respectively, the NE-1, NE-2, and NE-3 streamlines. The method employed to divide the maser spots of the NE region into three distinct streamlines is described in "Resolving the NE emission into three distinct streams" in the Supplementary Information of \citet{Mos22}. Colors denote the maser \Vlsr. The black, red, and blue dashed curves are the sinusoids corresponding to the plane-of-the-sky projection of the NE-1, NE-2, and NE-3 helical streamlines. The segments indicate the PA of the (sky-projected) direction of the magnetic field for the water masers detected in polarized emission (see Table~\ref{pol_fit}), using black or red color if the LOS inclination of the magnetic field is less or more than 55\degree, respectively.}
\label{NE_pol}
\end{figure}

\begin{figure}%
\includegraphics[width=0.5\textwidth]{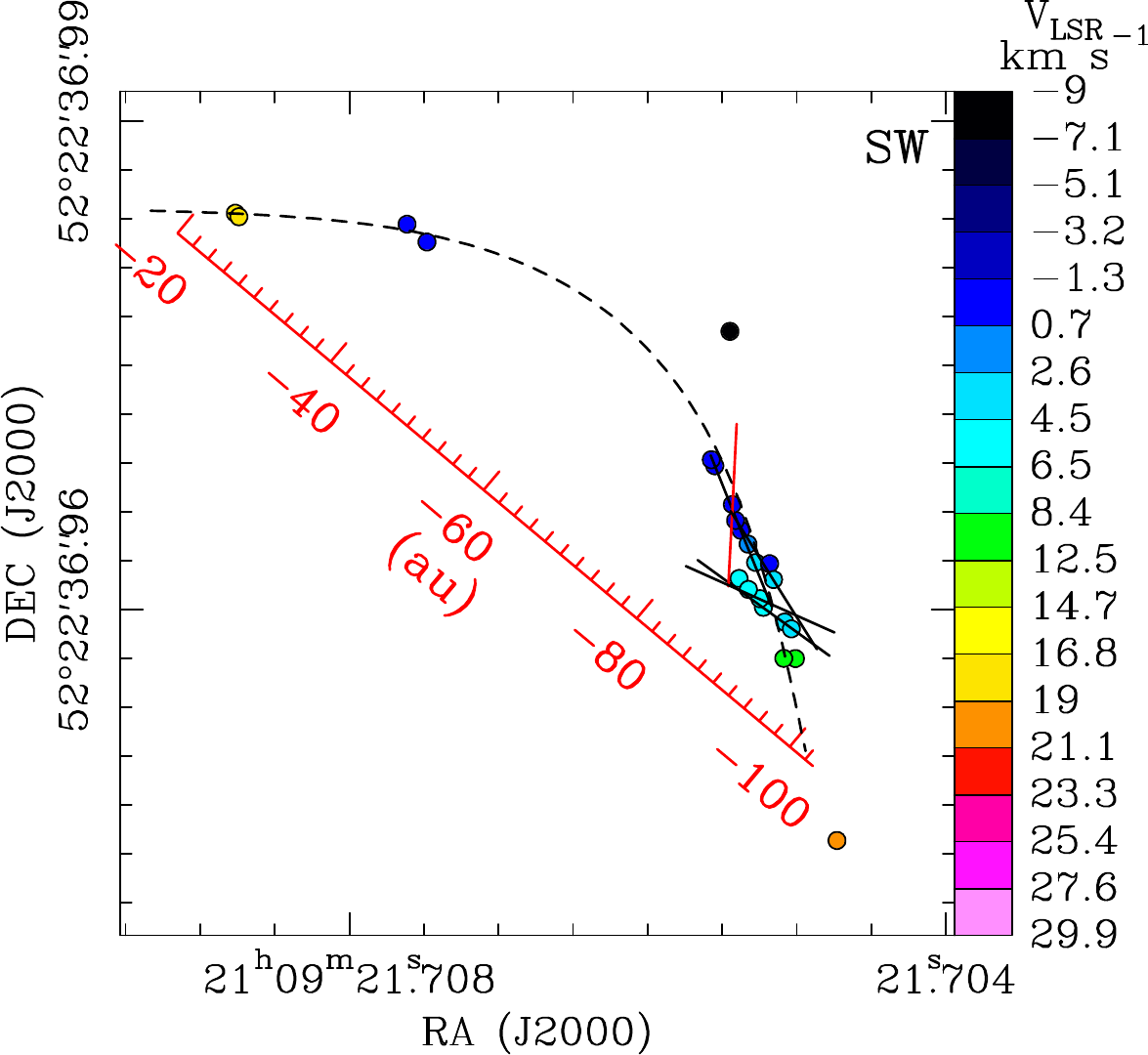}
\caption{{\bf Magnetic field orientation in the SW region.} Colored dots, the black dashed curve, and the black and red segments have the same meaning as in Fig.~\ref{NE_pol}.}
\label{SW_pol}
\end{figure}

\begin{figure}%
\includegraphics[width=0.5\textwidth]{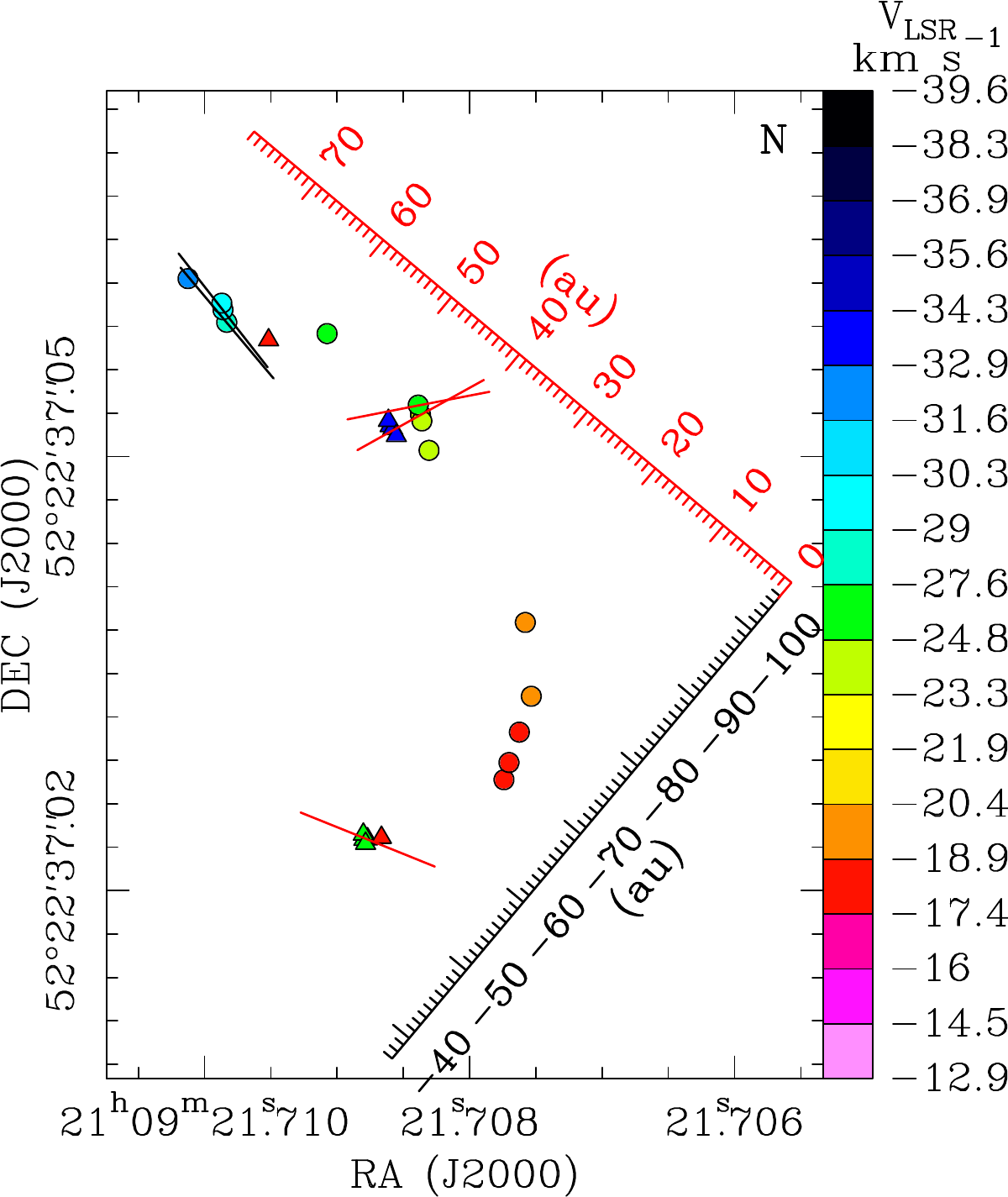}
\caption{{\bf  Magnetic field orientation in the N region.} Colored dots and triangles give the absolute positions of the 22~GHz water masers, with colors denoting the maser \Vlsr. The triangles mark a few masers with detached \Vlsr\ that belong to a streamline different from that traced by the dot-marked masers. The black and red segments have the same meaning as in Fig.~\ref{NE_pol}.}
\label{N_pol}
\end{figure}

\section{Results}
\label{Res}

\subsection{Maser polarization and 3D magnetic field geometry}

\label{Res-1}

Table~\ref{pol_fit} lists the water masers towards \targ\ \citep [see Supplementary Table~1 in][]{Mos22} with detected linear and/or circular polarizations. 
A weak level of linear polarization in the range of\ 0.3\%--3.2\% is observed in 24 masers, while 8 masers present a weak fraction of circular polarization of\ 0.4\%--1.8\%. All the masers with detected polarized emission have intensities of \ $\ge$0.7~\Jyb. While the PA of the linear polarization is accurately determined, with errors of mostly $\le$~10\degree, the LOS inclination, $\theta$, of the magnetic field has a large uncertainty, typically of several tens of degrees (see Table~\ref{pol_fit}). 

Our knowledge of the maser kinematics, together with the simulations,  provides an independent check on the 3D geometry of the magnetic field. 
One notable result is that, in the co-rotating frame of the launch points, the streamlines and magnetic field lines coincide (see Sect.~\ref{simu-fit}), which is consistent with the predictions by \citet{Mos22}. Assuming that the field lines are the helical streamlines traced with the water masers, Fig.~\ref{NE_geo} plots the orientation (lower panel) and inclination (upper panel) of the magnetic field vector at the maser positions along the NE streamlines. As the helix angle, $\alpha_B$ (the angle with which the helical field line winds around the jet axis, see Fig.~\ref{hel-mot}), is defined by the relation \ $\tan(\alpha_B) = f_z \ \mathfrak{R} $ \citep[][Eq.~12]{Mos22}, where \ $f_z$ \ and $ \mathfrak{R}$ are the fitted spatial frequency and radius of the sky-projected maser sinusoid \citep[][see Table~1]{Mos22}, the only free parameter of the helix position is the inclination angle of the jet axis with the plane of the sky.
In Fig.~\ref{NE_geo}, we use a jet inclination of \ 15\degree\  at the center of the observationally constrained range of \ 0\degree--30\degree. We also verified that a change of 15\degree\ in the jet inclination produces a shift in the LOS magnetic field inclination of $\approx$10\degree. Accordingly, in the upper panel of Fig.~\ref{NE_geo}, LOS inclinations of \ $\ge$65\degree \ or \ $\le$45\degree\ can be safely taken above or below, respectively, $\theta_{\rm crit}$ = 55\degree, and the corresponding orientations of the magnetic field on the sky taken perpendicular or parallel, respectively, to the measured direction of the linear polarization. For LOS inclinations of the magnetic field falling in between \ 45\degree\ and 65\degree, which is always the case for the polarized masers of the SW streamline (not shown in Fig.~\ref{NE_geo}), the sky-projected orientation of the magnetic field is chosen in such a way as to fall closer to the value predicted by the simulations.  Figures~\ref{NE_pol}~and~\ref{SW_pol} show the derived orientation of the magnetic field on the plane of the sky along the NE and SW streams, respectively. We stress that the resolution of the 90\degree\ ambiguity for the orientation of the magnetic field based on maser kinematics in conjunction with simulations is highly consistent with the results from the maser polarization fit. There is agreement for 15 out of the 18 masers with polarized emission across the NE and SW regions. For the three remaining masers, the predicted probability that the magnetic field is parallel or perpendicular to the linear polarization is marginal, that is, 63\%, 43\%, and 64\% for the maser labels \# 7, 18, and 30 (see Table~\ref{pol_fit}).

The masers in the N region are found at larger radii ($\ge$50~au), where our simulations do not properly reproduce the maser kinematics (see Sect.~\ref{simu-fit}), which prevents us from using the simulation results to constrain the inclination of the magnetic field. Figure~\ref{N_pol} plots the magnetic field orientations as predicted by the maser polarization fit. The magnetic field associated with the maser of label \#3, with only 53\% probability of being oriented parallel to the linear polarization, has been flipped by 90\degree\ to agree with the orientation of the nearby maser \#12, which shows 100\% probability of being perpendicular to the linear polarization (see Table~\ref{pol_fit}).

\subsection{Simulations}
\label{simu}


We attempted to reproduce the gas kinematics and magnetic field traced by the water masers using ab initio simulations to model a disk--jet system during the formation of a massive star. We previously used one of the simulations of our catalog for an order-of-magnitude comparison of maser kinematics in \citet{Mos22}. The reader can find an in-depth discussion about the methods and dynamics of the magnetically driven outflows and accretion disk formed during the simulations in \citet{Oli23a, Oli23b}.

 
Our ab initio simulations use the MHD methods to model the gravitational collapse of the weakly ionized gas and dust in a cloud core with the code Pluto \citep{Mig07}. Ohmic resistivity is considered as a nonideal MHD effect by using the model of \citet{Mac07}. We included additional modules for self-gravity \citep{Kui10} and the transport of the thermal radiation emitted by the gas and dust \citep{Kui20}.
 

The cloud cores used have initial masses of 100~\ms\ and radii of  0.1~pc. Each simulation in the catalog corresponds to a different initial condition for the gravitational collapse in an attempt to model the specific prestellar environment of \targ. Section~\ref{discu-ini} discusses our choice for the grid of initial conditions and the lesson learned from the simulations that deviate significantly from the observations. Here, we focus on the simulation that best reproduces the maser data.


We compared the maser data to several instants in the time evolution of different simulations from the catalog. The simulation that best reproduces the water masers, as discussed in Sect.~\ref{simu-fit}, assumes the following conditions at the onset of gravitational collapse ($t=0$): the density is distributed according to $\rho \propto r^{-1.65}$, and the cloud core rotates differentially according to angular velocity $\Omega \propto R^{-0.825}$ and with a rotational energy equivalent to 2\% of its gravitational energy content\footnote{In this context, $r$ is the spherical radius and $R$ the cylindrical radius.}. The magnetic field is assumed to be initially uniform and oriented parallel to the rotation axis. Its strength is determined by the mass-to-flux ratio, which we take as 20 times the critical (collapse-preventing) value \citep{Mou76}. A constant value of the opacity of $1\,\mathrm{cm}^2\,\mathrm{g}^{-1}$ was used to model the gas and dust, using an initial dust-to-gas mass ratio of 1\%. For this simulation, we used a grid of 448 cells logarithmically spaced in the spherical-radial direction and 80 cells linearly spaced in the polar direction, assuming axial and equatorial symmetry. This simulation was not part of the catalog in \citet{Oli23b}; it was a result of a refinement of parameters to model \targ. 


As described in \citet{Oli23a, Oli23b}, an accretion disk is formed after the initial gravitational collapse, when sufficient angular momentum is transferred toward the center of the cloud core. 
The magnetic field lines are dragged to the center of the cloud core by gravity, and are wound by rotation. Eventually, enough magnetic pressure builds up at the center to overcome gravity, forming a region of  low density (cavity) and thrusting a bow shock outward in the process.
Once the cavity is formed, the Alfvén speed increases in its interior because of the low density and the strong magnetic fields close to the protostar. The flow then becomes sub-Alfvénic, that is, the gas flow is forced to follow the magnetic field lines thanks to the restoring magnetic forces.
 When this happens, the magneto-centrifugal mechanism launches a fast jet (speeds $\gtrsim 100$~\kms) in a way consistent with the literature \citep[e.g.,][]{Bla82, Koe18}. The jet is narrow (typical radius of 50~au) and the flow becomes helical in motion around the regions that correspond well to the positions of the observed masers in the NE and SW regions, as we previously noted in \citet{Mos22}. The simulation data also reveal a slower and broader tower flow driven by magnetic pressure with typical speeds of\ 10~\kms. 
This magnetic pressure gradient originates in the winding of magnetic field lines on top of the disk as a result of rotation.
The tower flow broadens with time and is present at large scales (thousands to tens of thousands of astronomical units away from the YSO). This flow is consistent with the morphology and kinematics traced by the emission of the SO $J_N =  6_5-5_4$ line reported in \citet{Mos21}.

We searched the simulations in the catalog for instants where the main features of the disk and jet simultaneously satisfy the observational constraints for \targ\ given in \citet{Mos21}. The mass of the YSO should be within the interval \ 5.6$\pm$2~\ms. The accretion disk should have a radius of the order of\ 200~au. A bow shock was observed in \targ\ at a distance of $\approx 36000\,\mathrm{au}$ away from the YSO, giving a useful guess for the time evolution of the system. Finally, the magnetic field strength in the inner jet cavity in the simulation should lie within the estimates obtained with the water maser polarization data.

\begin{figure*}
\includegraphics[width=0.75\textwidth]{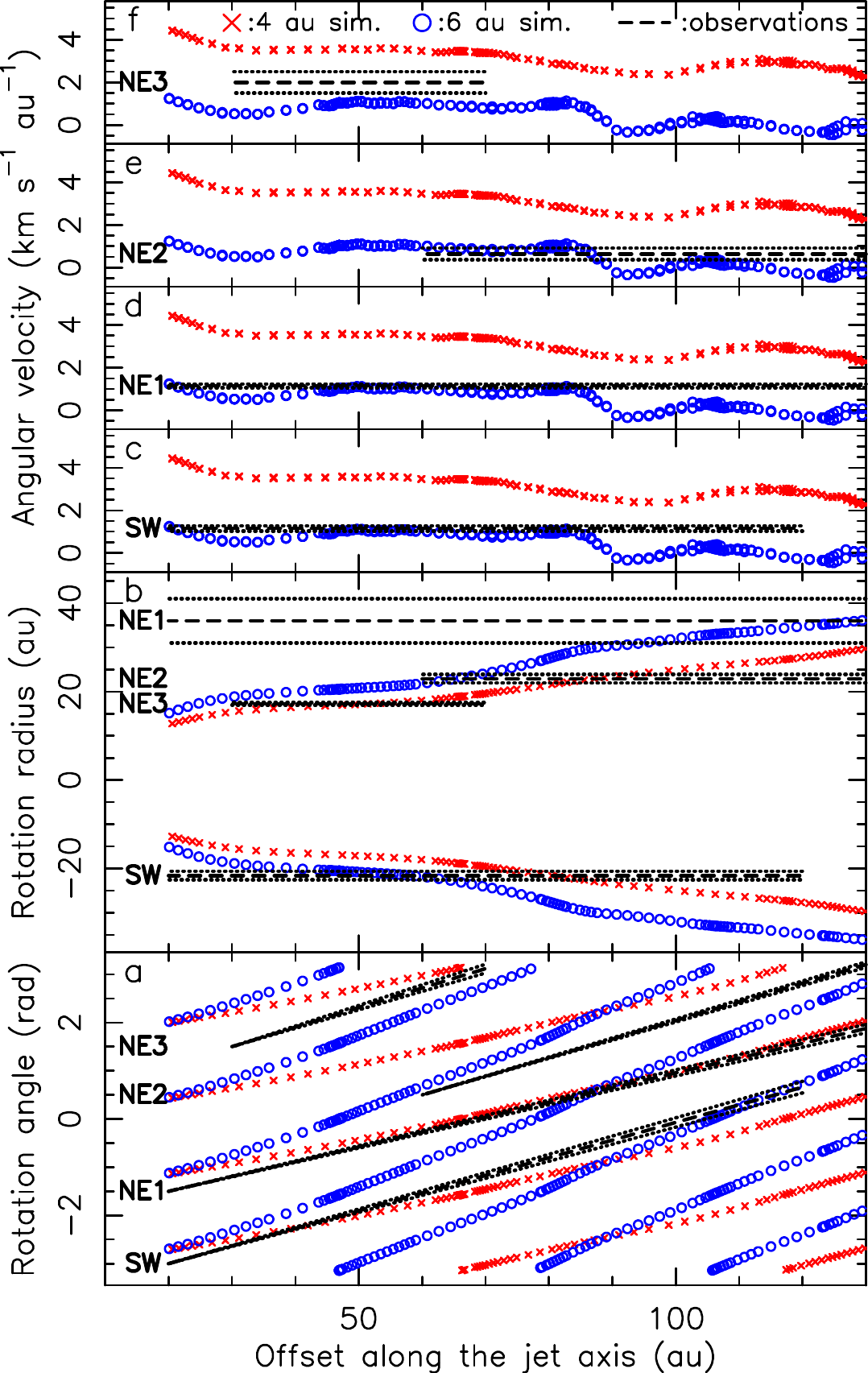}
\caption{{\bf Kinematics of the masers in comparison between the observations and the simulated streamlines.}
In all panels, red crosses and blue circles refer to the simulated streamlines emerging from the launch radii of 4 and 6~au, respectively, while the black dashed and dotted lines report measurements and corresponding errors for the four labeled maser streams \citep[][see Table~1]{Mos22}. For each maser stream, the measurements are reported over a range of distances along the jet axis matching the extension of the stream in the observations \citep[][see Figs.~3~and~4]{Mos22}.
\ Panel~(a):~Plot of the rotation angle versus the offset along the jet axis, which is taken positive also for the SW stream. A set of simulated streamlines launched at different azimuthal angles is shown. \ Panel~(b):~Plot of the rotation radius versus the offset along the jet axis. We note that the rotation radius for the SW streamline is taken to be negative. \  Panels~(c), (d), (e), and (f):~Plots of the angular velocity versus the offset along the jet axis, comparing the simulations with the SW, NE-1, NE-2, and NE-3 streams, respectively.}
\label{Kin_obs-sim}
\end{figure*}

\begin{figure}%
\includegraphics[width=0.5\textwidth]{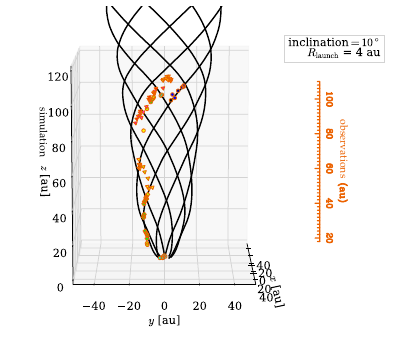}
\includegraphics[width=0.5\textwidth]{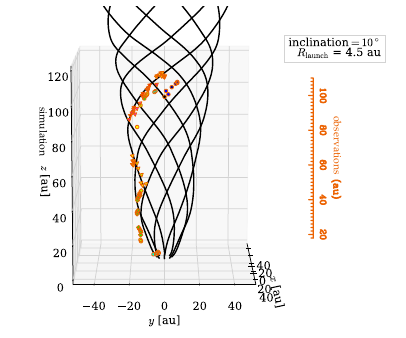}
\caption{{\bf Observations compared with simulations for the streamlines in the NE region.} \ Dots, triangles, and squares have the same meaning as in Fig.~\ref{NE_pol}. The black curves are the plane-of-the-sky projection of the simulated DW streamlines for the masers in the NE region. Two slightly different launch radii of 4 (upper panel) and 4.5~au (lower panel) are used to reproduce the NE-3 (squares) and NE-2 (triangles) streams, respectively. We note that the observations are rotated 50\degree \ clockwise so as to align the jet axis with the vertical axis.
}
\label{NE_obs-sim}
\end{figure}

\begin{figure}
\includegraphics[width=0.5\textwidth]{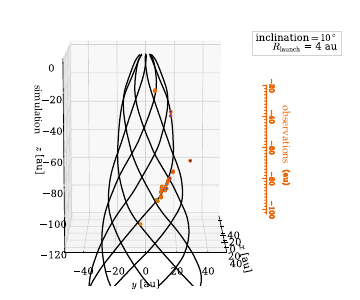}
\caption{{\bf Observations compared with simulations for the streamline in the SW region.} \ Dots have the same meaning as in Fig.~\ref{SW_pol}. The black curves are the plane-of-the-sky projection of the simulated DW streamlines in the SW region.  We note that the observations are rotated 50\degree \ clockwise so as to align the jet axis with the vertical axis.
}
\label{SW_obs-sim}
\end{figure}

\begin{figure*}
\includegraphics[width=\textwidth]{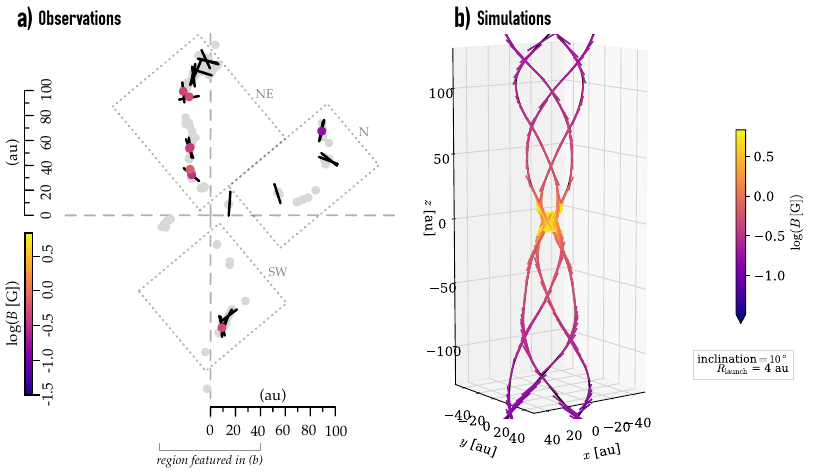}
\caption{{\bf Observations compared with simulations for the magnetic field in the NE and SW regions.} 
 Left~panel:~Dots give the water maser positions, using colors (gray) if the magnetic field amplitude has (not) been measured. Black segments show the direction of the magnetic field vector evaluated in correspondence with the masers with detected linear polarization. We note that the observed positions are rotated 50\degree \ clockwise so as to align the jet axis with the vertical axis. \ Right~panel:~3D view of four simulated streamlines departing from the launch radius of 4~au at 90\degree\ azimuthal separation from one another. The simulation snapshot refers only to the inner jet cavity, within a radius of 40~au. Segments show the direction of the magnetic field vector evaluated at different positions along the streamlines, and the colors indicate the decimal logarithm of the magnetic field intensity.
}
\label{B_obs-sim}
\end{figure*}

Once candidate simulation snapshots were chosen, we integrated streamlines for the vector field $\vec v - \omega_K R \, \vec{e}_\phi$, where \ $R$ \ is the cylindrical radius, $\vec{e}_\phi$ \ is the unit vector along the azimuthal direction, and \ $\omega_K$ \ is the (quasi-)Keplerian angular velocity of a given launching point located at $R_\text{launch}, z_\text{launch}$. We checked that those streamlines coincide with the magnetic field lines within the sub-Alfvénic regime, as expected from theory. The positions of the launching points were chosen by trial and error, taking the expected values from the kinematic analysis (sinusoidal fit) in \citet{Mos22} as a guide. We computed the velocity along the LOS, respecting the observational constraint of $0^\circ$--$30^\circ$ for the inclination angle of the jet with respect to the plane of the sky.

\subsection{Reproduction of the maser kinematics}
\label{simu-fit}

We searched for the streamlines that best reproduce the maser sinusoids in the NE and SW regions over the range of launch radii \ 2--20~au \ constrained by the observations \citep[][see Table~2]{Mos22}. At these radii, the simulated streamlines are approximately helixes and their shapes are defined by the rotation radius \ $ \mathfrak{R}$ \ and the helix angle \ $\alpha_B$ (see Fig.~\ref{hel-mot}).  For a given helix shape, the helical motion depends only on\ $V_z$, the velocity component along the jet axis, because the angular velocity is determined by the relation \ $\omega_{\rm e} = \tan(\alpha_B) \ V_z \ / \ \mathfrak{R}$. The angular velocity \ $\omega$ \ measured for the maser streamlines in the NE and SW regions \citep[][see Table~1]{Mos22} can be calculated with the equation \ $\omega = \omega_{\rm K} - \omega_{\rm e} $ \citep[][Eq.~6]{Mos22}, where \ $\omega_{\rm K}$ \ is the (quasi-)Keplerian angular velocity of the launch point from simulations, and the minus sign accounts for the different sense of rotation of the maser streamlines between the reference frame of the observer and that co-rotating with the disk.

The chosen simulation snapshot for modeling the maser data using the initial conditions described in Sect.~\ref{simu} is at $11.2\,\mathrm{kyr}$ of evolution. At that time, the YSO has a mass of $5.0\,M_\odot$, and the accretion disk has a radius of about $255\,\mathrm{au}$. 
We computed the average speed of propagation of the bow shock across the computational domain of \ $0.1\mathrm{\,pc} = 20\, 626\mathrm{\,au}$, obtaining around $6500\mathrm{\,km\,s^{-1}}$. Extrapolating to the reference time $t=11.2\mathrm{\,kyr}$, we estimate that the bow shock will have propagated up to $\sim 33\,000\,\mathrm{au}$, which is close to the observed value of $36\,000\,\mathrm{\,au}$. 

By considering both the geometry and kinematics, we verified that only the streamlines from launch radii \ $\mathfrak{R}_{\rm K}$ \ in the narrow range of\ 4--6~au \ can be consistent with the maser observations. For these launch radii, the simulated streamlines  over the maser region yield values of \ $\mathfrak{R}$ varying in the range of\ 15--35~au, $f_z$ in the range \ 0.03--0.05~rad~au$^{-1}$, and  angular velocity \ $\omega$ ranging from \ 0 to 4 \kms\ au$^{-1}$, which are all highly consistent with our previous measurements \citep[][see Table~1]{Mos22}. This is illustrated in Fig.~\ref{Kin_obs-sim}. The linear change of the rotation angle $\phi$ with the distance $z$ along the jet indicates that the simulated streamlines are helixes, as observed for the maser streams; in addition, the values of $ f_z = {\rm d}\phi / \rm{d} z $ are in very good agreement between observations and simulations. The rotation radius and angular velocity of the simulated streamlines are also consistent with the measurements for the four maser streams.
Figures~\ref{NE_obs-sim}~and~\ref{SW_obs-sim} present the overlay of the maser positions and simulated streamlines in the NE and SW regions, respectively. For the NE-1, NE-2, NE-3, and SW streams, the rms deviations of the separation between the maser and the simulated positions are 5.4, 1.3, 0.6, and 1.1 au, respectively. These small deviations prove  the degree of consistency of our simulations with the observations.  Figure~\ref{B_obs-sim} compares the observed distribution of magnetic field orientation and amplitude with the simulated field lines.


By comparing Figs.~\ref{glo}b,~\ref{NE_obs-sim},~and~\ref{SW_obs-sim}, we see the same kinematic patterns in $V_\mathrm{LSR}$ present in both the simulations and the maser data. Label A in Fig.~\ref{glo}b shows a region where $-10 \lesssim V_\mathrm{LSR} \lesssim 20 \,\mathrm{km\,s^{-1}}$, similarly to what has been observed in the NE-3 streamline. In both cases, the streamline is located within a vertical distance of 20 to 60 au away from the YSO, although this is easily adjustable in the simulations by changing the azimuthal angle of the launching point. Streamline NE-2 extends from \ $50 \lesssim z \lesssim 120\,\mathrm{au}$, and is better reproduced by considering a slightly larger launching point (see Fig.~\ref{NE_obs-sim}, for $R_\text{launch}=4.5\,\mathrm{au}$) or a slightly larger inclination. In NE-2, $V_\text{LSR}$ changes from positive to negative for $  100 \lesssim z \lesssim 120\,\mathrm{au}$. The same pattern is obtained in the simulations, as marked by label B 
in Fig.~\ref{glo}b. Finally, the  $V_\text{LSR}$  pattern in the streamline labeled C in Fig.~\ref{glo}b (negative to positive values) is analogous to the one observed in the SW region for $ 80 \lesssim z \lesssim 100$ (see also Fig.~\ref{SW_obs-sim}).


The simulation data reveal that the outflow is launched from different radii across the disk, even at distances of $R =$ 100~au or more. However, it is not launched through the magneto-centrifugal mechanism but it belongs to the magnetic-pressure driven tower flow instead. We discuss the comparison of wider streamlines with the N region and the role of ambipolar diffusion (which is missing from our current setup) in Sect.~\ref{discu}.

\section{Discussion}
\label{discu}

\subsection{Our best-matching simulation}


\begin{figure}
        \centering
        \includegraphics[width=\linewidth]{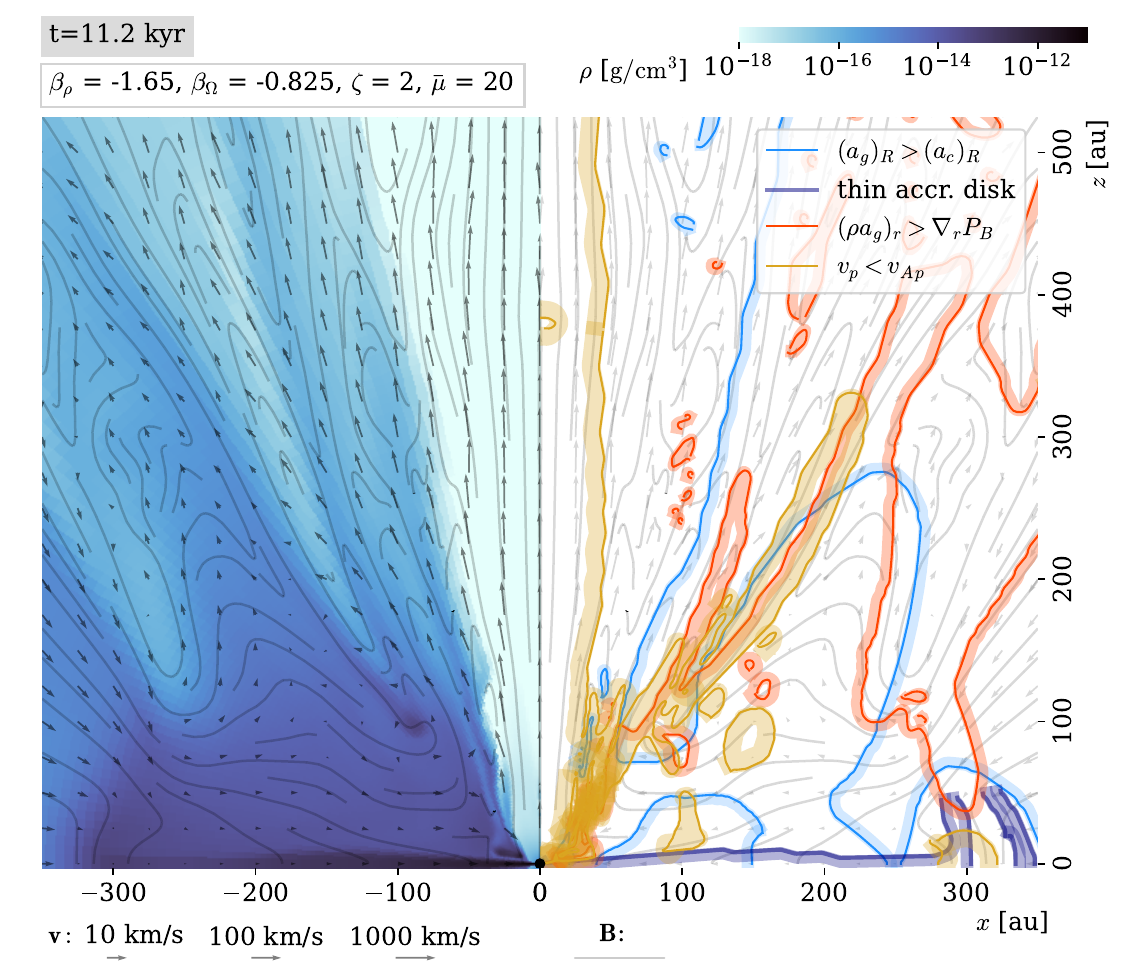}
        \caption{{\bf Morphology and dynamics of the jet and accretion disk for the reference simulation used to reproduce the water maser data.}  Left~panel:~Density, velocity, and magnetic field lines of a vertical slice of the simulation. Scales for each physical quantity are shown. \ Right~panel:~Shadowed contours that indicate the ratio of two dynamically relevant quantities. Each contour is composed of a solid line and a shadow of the same color. The shadowed part indicates the region where the corresponding inequality listed in the legend holds true.}
        \label{fig:split}
\end{figure}

An overview of the morphological and dynamical structure of the disk--jet system computed with the reference simulation is presented in Fig.~\ref{fig:split}. The lowest-density region constitutes the narrow cavity where the magneto-centrifugal jet is launched and accelerated. The launching region of the jet can be seen as the intersection of two regions determined by the contours on the right panel: the shadowed part of the yellow contour and the unshadowed part of the blue contour. These two regions cover the area where the flow is sub-Alfvénic (the former) and simultaneously where the centrifugal force dominates over gravity (the latter), forcing material from the inner disk to be expelled following the magnetic field lines.

The accretion disk consists of two layers: a thin layer supported by thermal pressure, and another layer supported by magnetic pressure. The radius of the disk is about $255\,\mathrm{au}$ if only counting the section that is rotating with Keplerian-like speeds; a centrifugal barrier exists between the outer edge of the disk and the infalling envelope. As seen from the orange contour, the magnetic pressure gradient is larger than gravity on top of the disk, and this drives the flow at cylindrical distances of $\gtrsim 50\,\mathrm{au}$ away from the YSO.

Figures~\ref{NE_obs-sim}--\ref{B_obs-sim} show that our simulations are able to closely reproduce the geometry and kinematics of the water maser streamlines for the spiral motions in the NE and SW regions. These plots also demonstrate that the magnetic field lines coincide with the streamlines; that is, the magnetic field vectors are tangent to the streamlines at any point. 
Along the NE and SW spiral motions, the measured sky-projected orientation of the magnetic field is, in reality, not tangent to the maser sinusoids (see Figs.~\ref{NE_pol}~and~\ref{SW_pol}), which is in contrast with the simulations. However, along the NE-2 and NE-3 streamlines, excluding three masers at the NE-2 extreme, the PA of the magnetic field remains consistently larger than that of the tangent direction by an approximately constant angle \ $\Delta \Theta = 27$\degree $\pm$ 8\degree.
Considering the SW streamline, the sky-projected field is directed close ---within 25\degree--- to the stream tangent in three of the five detected polarized masers, and is tilted by a larger angle ---\ 38\degree \ to  50\degree--- in the two remaining masers closer to the SW extreme. Toward \targ,\  a compact ionized wind has been observed in correspondence with the water maser region \citep{Mos16,Mos21}. In the following, we consider the case where the maser emission arises from weakly ionized gas, where Faraday rotation can rotate the polarization angle of the maser emission.

From the expression for optically thin emission of ionized gas  \citep[][see Eq.~10.36]{Wil00}, with an electron temperature of\ 10$^4$~K and using the peak brightness temperature of \ $\approx$200~K of the VLA A-array K-band continuum observed by \citet{Mos16}, we derive an emission measure of \ $\sim$4 $\times$ 10$^{7}$~pc~cm$^{-6}$.
As the water masers move along spirals with radii of \ $\approx$20~au \citep[][see Table~1]{Mos22}, and the ionized core of the jet should be found inside the external shell of (predominantly) molecular material, we employ a size for the ionized jet of 40~au. From these  deductions and the above estimate of emission measure, the derived value for the electron density is \ $\sim$5 $\times$10$^5$~cm$^{-3}$. This is in good agreement with the lowest densities inside the flow cavity from our simulations (see Fig.~9).

\begin{figure*}
        \includegraphics[width=\textwidth]{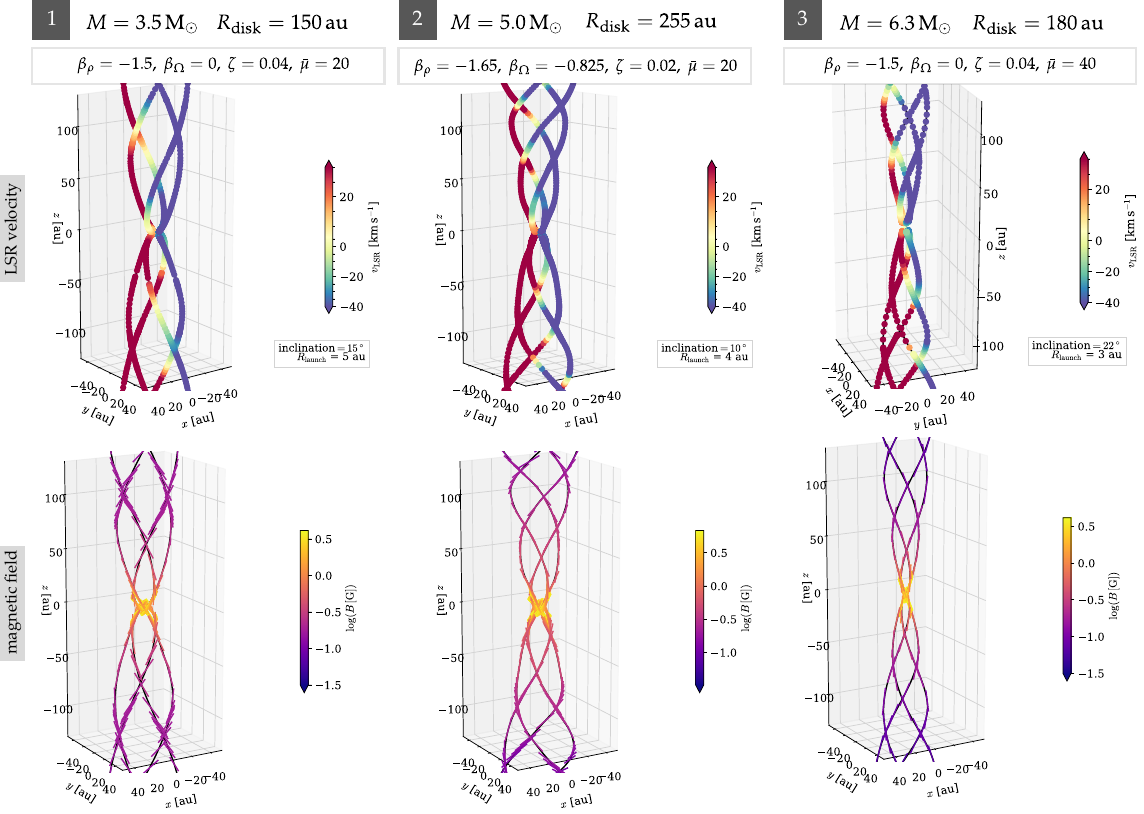}
        \caption{{\bf Comparison of streamlines produced by different simulations in the catalog.} The simulation described in Sect.~\ref{simu} corresponds to column 2. The initial conditions in each case are described by the following parameters: $\beta_\rho$ is the exponent of the density profile, $\beta_\Omega$ is the exponent of the angular velocity profile, $\zeta$ is the ratio of rotational to gravitational energy,  and $\bar \mu$ is the mass-to-flux ratio normalized to the critical (collapse-preventing) value.}
\label{comp-simu}
\end{figure*}

Comparing the fitted maser fluxes \ $T_{\rm{b}} \,\Delta\Omega$ \ in Table~\ref{pol_fit} with the threshold for water maser saturation of \ 9.83 log (K sr) \citep{Sur11b}, all the polarized masers detected in \targ\ (but features \# 1, 16 and 40) should be unsaturated. The detection of linear polarization from the water masers requires that Faraday rotation be small ---that is, < 1~rad--- over the maser unsaturated amplification length, which agrees with our finding that, in most cases, the sky-projected magnetic field presents a small deviation of $\le$~30\degree\  from the streamline tangent. Using the expression for Faraday rotation of \citet[][Eq.~5]{Fis06b}, an average value for the magnetic field of 430~mG (see Table~\ref{pol_fit}), and an unsaturated amplification path of \ $\sim$~1~au taken equal to the sky-projected maser size \citep[][see Supplementary Fig.~6]{Mos22}, the average tilt from the tangent direction of \ 27\degree\ observed in the NE stream would correspond to an electron number density of \ 2 $\times$ 10$^3$~cm$^{-3}$. Comparing with the above estimate for the ionized jet density, the ionization fraction of the surrounding molecular shell would be of \ $\sim$~10$^{-2}$.


 The magnetic field along the N streamline is poorly sampled with only five masers at three separated positions (see Fig.~\ref{N_pol}). The orientation of the magnetic field is approximately perpendicular to the streamline tangent at heights of \ $\approx$10~and~40~au and is almost parallel to the tangent at \ $\approx$~70~au. 
We consider it unlikely that such a large change in the orientation of the magnetic field is due to Faraday rotation.
Indeed, the N streamline is found at significantly larger radii, 50--100~au, than the NE and SW streamlines, and one might expect the ionization fraction to be  negligible  there, which would mean no observable amount of Faraday rotation.

Despite the large errors, our measurements of the magnetic field amplitude are all within a relatively small range of \ 100--700~mG (see Table~\ref{pol_fit}), which is in good agreement with the simulations (see Fig.~\ref{B_obs-sim}).
Such a high value of the magnetic field might result from the shock-ionization of the jet cavity, which reduces the magnetic (ambipolar, Hall, Ohmic) diffusivity. Recent models that also include ambipolar diffusion but neglect the effects of ionization in the jet cavity find a stringent cap for the magnetic field at a few 100~mG \citep{Com22}. Our model also does not include shock ionization, but because Ohmic resistivities are low for low densities, such as the ones in the jet cavity, we should be able to more accurately reproduce the observed magnetic field strengths. However, a more realistic description of the problem would require that we consider the combined effects of shock-ionization and ambipolar diffusion.

In particular, ambipolar diffusion can be crucial for reproducing the right thickness of the disk at  radii $\ge$~50~au, which in our simulations is probably inflated by locally excessive magnetic pressure. We expect that reducing the magnetic pressure would make the disk flatter and lower the density on top of the disk. If the density, $\rho$, is low enough such that the Alfvén velocity $v_{\rm A} = B/\sqrt{4\pi\rho}$ increases up to the point where the flow becomes sub-Alfvénic, the conditions for the magneto-centrifugal mechanism would be satisfied. 
This means that streamlines launched from the top of the disk at radii $\ge$50~au might better match the observed maser kinematics in the N region.
In conclusion, the incorporation of ambipolar diffusion could be crucial for setting the conditions for the magneto-centrifugal launch of the flow at large radii, and we plan to address this important issue in future works.

\subsection{Exploring the initial conditions}

\label{discu-ini}

As mentioned in Sect.~\ref{simu}, the simulation catalog explores different initial density and angular momentum distributions, as well as different initial magnetic field strengths. We did not explore different masses of the cloud core because of the results obtained in \citet{Oli23a,Oli23b}: different cloud core masses produce results that are approximately scalable with the free-fall time within a mass interval of at least 50--150~\ms. With this in mind, we searched the simulation catalog for examples of simulations that approximately reproduce the mass of the protostar, the size of the disk, and the kinematics of the flow for a given time, following the analysis by \citet{Oli23a}. Figure~\ref{comp-simu} shows streamlines with \Vlsr\ and magnetic field from two additional simulation snapshots (labeled as 1 and 3) at each side of the reference simulation discussed in Sect.\ref{simu} (labeled as 2). 

In comparison to the reference simulation~2, simulations~1~and~3 start from a slightly flatter density profile. Both simulations~1~and~3 start from solid body rotation instead of differential rotation (the latter choice concentrates angular momentum at the center of the cloud core at earlier times), but there is less rotational energy content in the cloud core of simulation~2. Simulation~3 considers a weaker initial magnetic field than the rest. 

All three simulations produce streamlines with similar helical shapes at different points in time. With this choice, we intend to exemplify three YSO masses within the interval constrained by previous   observations, that is, 5.6$\pm$2~\ms: snapshot~1 considers a YSO mass of $3.5\,M_\odot$, which is on the lowest part of the mass interval; snapshot~2, a mass of $5.0\,M_\odot$, which is around the middle of the interval; and snapshot~3, a mass of $6.3\,M_\odot$, which is above the midpoint of the interval. Because of this, the launching points chosen for approximately matching the \Vlsr\ to the maser data are different in every case. This means that the helix is wider for simulation~1, and narrower for simulation~3. All three helical patterns match the maser velocity distribution within an order of magnitude. 

The bottom half of Fig.~\ref{comp-simu} displays the direction of the magnetic field (line segments) and its strength (color scale). The magnitude of the magnetic field in all cases is within $0.1$ and $1\,\mathrm{G}$ for the regions where the masers are found. The fact that the magnetic field and the flow streamlines are parallel in all cases shows that we are dealing with a sub-Alfvénic flow.

By finding several examples of simulations that model the maser data, we aimed to constrain the initial conditions for massive star formation, which is a difficult problem due to the large amount of parameters involved. Even though we find several simulation snapshots and launching points that reproduce the observations, there are many combinations of initial conditions, launching points, and inclinations for which the helical pattern does not fit the constraints in dimensions (e.g., a relatively constant helix radius as a function of height) or the observed  \Vlsr. For example, according to Fig.~9 of \citet{Oli23a}, an initial steep density profile of $\rho \propto r^{-2}$ produces a YSO of $10\,M_\odot$ when the disk is $\sim 100\,\mathrm{au}$ in radius, which does not satisfy the observational constraints. An additional complication comes from the fact that the simulations show episodic ejections of material that temporarily alter the shape of the cavity, and therefore the streamlines (we chose snapshots that are as free from transient phenomena as possible). We note that a few cases of water maser flares have recently been associated to episodes of accretion bursts in high-mass YSOs \citep[see, e.g.,][]{Bro18}. However, the water masers observed in \targ\ are not flaring and are probably tracing a quiescent phase of the YSO's accretion or ejection process.

Even though our scan of all possible initial conditions is incomplete, our comparison shows that there are initial conditions that lead to  efficient reproduction of the flow observed in \targ. Surveys of cloud cores in early stages of massive star formation \citep[see, e.g.,][]{Gie22,Beu18} have found density profiles with power-law exponents between $-$1.5 and $-$2.6. The results of our analysis favor flatter density profiles within this range. The initial angular momentum is consistent with the results from \citet{Good93}, that is, a rotational energy content of a few percent of the gravitational energy. Finally, our results favor high mass-to-flux ratios, which mean magnetic field strengths of the order of tens of micro Gauss, or only one order of magnitude larger than what it is expected for the initial conditions in low-mass star formation. This last result should be taken with caution, as it might change upon the inclusion of ambipolar diffusion in the models.

\section{Conclusions}

\label{conclu}

We recently performed  global VLBI polarimetric observations of the 22~GHz water masers toward \targ. 
In this source, we have previously used the maser emission to trace the kinematics of a MHD DW, and here we complement the previous work with information about the maser polarization in order to estimate the underlying magnetic field.
We compared our observations of gas kinematics and magnetic fields in \targ\ with resistive-radiative-gravito-MHD simulations of the formation of a massive star starting from the gravitational collapse of a rotating cloud core (radius of 0.1~pc, mass 100~\ms), with a centrally dominated density distribution ($\rho \propto r^{-1.65}$), in differential rotation, and threaded by a uniform magnetic field.  We explored a range of initial conditions, parametrized by the exponents of the density and angular velocity power laws, the ratios of rotational to gravitational energy, and the mass-to-magnetic-flux ratios, searching for simulation snapshots that better reproduce our observational results. After the initial search, we were able to refine the parameter space and run a simulation that closely matches the observed paths of the masers, their \Vlsr, and the magnetic field strengths measured with polarization.

The achieved sub-mJy sensitivity allows us to significantly increase the number of detections of polarized masers from
the typical figure of a few to a few tens, and to sample the magnetic field over radii of \ $\le$~100~au from the YSO. All the masers with detected polarization have intensities of $\ge$0.7~\Jyb\ and are among the 40 strongest ones. A set of 24 masers present a weak level of linear polarization in the range of 0.3\%--3.2\%, and 8 masers show a weak fraction of circular polarization of \ 0.4\%--1.8\%. While the PA of the linear polarization is directly measured with typical errors of\ $\le$~10\degree, the model-derived LOS inclination of the magnetic field is in most cases loosely constrained with uncertainties of several 10\degree. We further constrained the 3D geometry of the magnetic field by considering two main findings: (1)~the magnetic field lines should coincide with the flow streamlines, as clearly pointed out by the simulations; and (2)~the 3D helical geometry of the streamlines, as traced by the water masers. Close to the jet axis ($\le$~25~au), all along the three best-sampled streamlines, the orientation of the magnetic field in the plane of the sky is found to deviate from the streamline tangent in most cases by $\le$~30\degree. Along the stream at larger radii (50--100~au), the magnetic field is measured at only three separated positions, and its orientation is found to be approximately perpendicular to the streamline tangent at heights of \ $\approx$10~and~40~au, and parallel to the tangent at \ $\approx$70~au. The small tilt in the magnetic field direction observed along the inner streams is consistent with Faraday rotation, assuming that the molecular shell of the jet has a low\  level of ionization, of namely $\sim$~10$^{-2}$. Despite the large errors, the values of magnetic field amplitude inferred from maser circular polarization are all within a relatively small range of \ 100--700~mG, which is in good agreement with the simulation results and consistent with reduced magnetic diffusivity in the jet cavity owing to efficient shock ionization. 

We searched for additional candidates  within our simulation catalog able to reproduce the observational results  in order to constrain the initial conditions for the gravitational collapse of a cloud core for massive star formation. Although we find more than one viable candidate set of initial conditions, we find agreement with typical values of density distribution and the  ratio of rotational to gravitational energy in dense cores observed in surveys of massive star forming regions.

 \section*{Acknowledgments}
 RK acknowledges financial support via the Heisenberg Research Grant funded by the German Research Foundation (DFG) under grant no.~KU 2849/9. The European VLBI Network is a joint facility of independent European,
    African, Asian, and North American radio astronomy institutes.
    Scientific results from data presented in this publication are 
    derived from the following EVN project code:  GM077.

%

%
%

\end{document}